\PassOptionsToPackage{unicode}{hyperref}
\PassOptionsToPackage{hyphens}{url}
\PassOptionsToPackage{dvipsnames,svgnames,x11names}{xcolor}
\documentclass[
  letterpaper,
  DIV=11,
  numbers=noendperiod]{scrartcl}
\usepackage{xcolor}
\usepackage{amsmath,amssymb}
\usepackage{iftex}
\ifPDFTeX
  \usepackage[T1]{fontenc}
  \usepackage[utf8]{inputenc}
  \usepackage{textcomp} % provide euro and other symbols
\else % if luatex or xetex
  \usepackage{unicode-math} % this also loads fontspec
  \defaultfontfeatures{Scale=MatchLowercase}
  \defaultfontfeatures[\rmfamily]{Ligatures=TeX,Scale=1}
\fi
\usepackage{lmodern}
\ifPDFTeX\else
\fi
\IfFileExists{upquote.sty}{\usepackage{upquote}}{}
\IfFileExists{microtype.sty}{% use microtype if available
  \usepackage[]{microtype}
  \UseMicrotypeSet[protrusion]{basicmath} % disable protrusion for tt fonts
}{}
\usepackage{setspace}
\makeatletter
\@ifundefined{KOMAClassName}{% if non-KOMA class
  \IfFileExists{parskip.sty}{%
    \usepackage{parskip}
  }{% else
    \setlength{\parindent}{0pt}
    \setlength{\parskip}{6pt plus 2pt minus 1pt}}
}{% if KOMA class
  \KOMAoptions{parskip=half}}
\makeatother
\makeatletter
\ifx\paragraph\undefined\else
  \let\oldparagraph\paragraph
  \renewcommand{\paragraph}{
    \@ifstar
      \xxxParagraphStar
      \xxxParagraphNoStar
  }
  \newcommand{\xxxParagraphStar}[1]{\oldparagraph*{#1}\mbox{}}
  \newcommand{\xxxParagraphNoStar}[1]{\oldparagraph{#1}\mbox{}}
\fi
\ifx\subparagraph\undefined\else
  \let\oldsubparagraph\subparagraph
  \renewcommand{\subparagraph}{
    \@ifstar
      \xxxSubParagraphStar
      \xxxSubParagraphNoStar
  }
  \newcommand{\xxxSubParagraphStar}[1]{\oldsubparagraph*{#1}\mbox{}}
  \newcommand{\xxxSubParagraphNoStar}[1]{\oldsubparagraph{#1}\mbox{}}
\fi
\makeatother

\usepackage{longtable,booktabs,array}
\usepackage{calc} % for calculating minipage widths
\usepackage{etoolbox}
\makeatletter
\patchcmd\longtable{\par}{\if@noskipsec\mbox{}\fi\par}{}{}
\makeatother
\IfFileExists{footnotehyper.sty}{\usepackage{footnotehyper}}{\usepackage{footnote}}
\makesavenoteenv{longtable}
\usepackage{graphicx}
\makeatletter
\newsavebox\pandoc@box
\newcommand*\pandocbounded[1]{% scales image to fit in text height/width
  \sbox\pandoc@box{#1}%
  \Gscale@div\@tempa{\textheight}{\dimexpr\ht\pandoc@box+\dp\pandoc@box\relax}%
  \Gscale@div\@tempb{\linewidth}{\wd\pandoc@box}%
  \ifdim\@tempb\p@<\@tempa\p@\let\@tempa\@tempb\fi% select the smaller of both
  \ifdim\@tempa\p@<\p@\scalebox{\@tempa}{\usebox\pandoc@box}%
  \else\usebox{\pandoc@box}%
  \fi%
}
\def\fps@figure{htbp}
\makeatother

\NewDocumentCommand\citeproctext{}{}

\makeatletter
 \let\@cite@ofmt\@firstofone
 \def\@biblabel#1{}
 \def\@cite#1#2{{#1\if@tempswa , #2\fi}}
\makeatother
\newlength{\cslhangindent}
\newlength{\csllabelwidth}
\newenvironment{CSLReferences}[2] % #1 hanging-indent, #2 entry-spacing
 {\begin{list}{}{%
  \setlength{\itemindent}{0pt}
  \setlength{\leftmargin}{0pt}
  \setlength{\parsep}{0pt}
  \ifodd #1
   \setlength{\leftmargin}{\cslhangindent}
   \setlength{\itemindent}{-1\cslhangindent}
  \fi
  \setlength{\itemsep}{#2\baselineskip}}}
 {\end{list}}
\usepackage{calc}

\providecommand{\tightlist}{%
  \setlength{\itemsep}{0pt}\setlength{\parskip}{0pt}}

\usepackage{booktabs}
\usepackage{longtable}
\usepackage{array}
\usepackage{multirow}
\usepackage{wrapfig}
\usepackage{float}
\usepackage{colortbl}
\usepackage{pdflscape}
\usepackage{tabu}
\usepackage{threeparttable}
\usepackage{threeparttablex}
\usepackage[normalem]{ulem}
\usepackage{makecell}
\usepackage{xcolor}
\KOMAoption{captions}{tableheading}
\makeatletter
\@ifpackageloaded{caption}{}{\usepackage{caption}}
\AtBeginDocument{%
\ifdefined\contentsname
  \renewcommand*\contentsname{Table of contents}
\else
  \newcommand\contentsname{Table of contents}
\fi
\ifdefined\listfigurename
  \renewcommand*\listfigurename{List of Figures}
\else
  \newcommand\listfigurename{List of Figures}
\fi
\ifdefined\listtablename
  \renewcommand*\listtablename{List of Tables}
\else
  \newcommand\listtablename{List of Tables}
\fi
\ifdefined\figurename
  \renewcommand*\figurename{Figure}
\else
  \newcommand\figurename{Figure}
\fi
\ifdefined\tablename
  \renewcommand*\tablename{Table}
\else
  \newcommand\tablename{Table}
\fi
}
\@ifpackageloaded{float}{}{\usepackage{float}}
\floatstyle{ruled}
\@ifundefined{c@chapter}{\newfloat{codelisting}{h}{lop}}{\newfloat{codelisting}{h}{lop}[chapter]}
\floatname{codelisting}{Listing}

\makeatother
\makeatletter
\@ifpackageloaded{caption}{}{\usepackage{caption}}
\@ifpackageloaded{subcaption}{}{\usepackage{subcaption}}
\makeatother
\usepackage{bookmark}
\IfFileExists{xurl.sty}{\usepackage{xurl}}{} % add URL line breaks if available
\makeatletter
\@ifundefined{xmpquote}{\newcommand{\xmpquote}[1]{#1}}{}
\makeatother
\hypersetup{
  pdftitle={Exploring Exchangeability within Permutation Inference},
  pdfauthor={Johanna Hardin; Lauren Quesada; Julie Ye; Nicholas J. Horton},
  pdfkeywords={\xmpquote{computational
methods}, \xmpquote{re-sampling}, \xmpquote{hypothesis
testing}, \xmpquote{clustered observations}},
  colorlinks=true,
  linkcolor={blue},
  filecolor={Maroon},
  citecolor={Blue},
  urlcolor={Blue},
  pdfcreator={LaTeX via pandoc}}

\title{Exploring Exchangeability within Permutation Inference}
\usepackage{etoolbox}
\makeatletter
\providecommand{\subtitle}[1]{% add subtitle to \maketitle
  \apptocmd{\@title}{\par {\large #1 \par}}{}{}
}
\makeatother
\subtitle{Implications for Statistics and Data Science Educators}
\author{Johanna Hardin \and Lauren Quesada \and Julie Ye \and Nicholas
J. Horton}
\date{2026-09-26}
\begin{document}
\maketitle
\begin{abstract}
Permutation tests are a powerful and flexible approach to inference via
resampling. As computational methods become more ubiquitous in the
statistics curriculum, use of permutation tests has become more
tractable. At the heart of the permutation approach is the
\emph{exchangeability condition}, which determines the appropriate null
sampling distribution. We explore the exchangeability condition in the
context of permutation tests, including settings where the
exchangeability condition is not tenable. Our examples include
two-sample comparisons of means, clustered designs, multiple linear
regression, and two-sample comparisons of correlation. The different
exchangeability challenges in each example allow further exploration of
the definition and importance of the exchangeability requirement for
permutation testing. We believe that understanding the exchangeability
condition and how it relates to the null hypothesis gives students the
skills to work effectively with permutation tests in more complicated
modeling settings and to further their understanding of statistical
inference. We close with pedagogical recommendations for instructors who
wish to incorporate permutation inference into their curriculum as a
means to increase student understanding of resampling-based inference.
\end{abstract}

\setstretch{2}
\newpage

\section{Introduction}\label{introduction}

Statistical inference attempts to infer properties of population
parameters from sample statistics. Many of the models and analysis
methods taught in a standard undergraduate statistics curriculum rely on
asymptotic normal theory, with the theoretical underpinnings driven by
the Central Limit Theorem. However, permutation tests, a form of
resampling-based inference,\footnote{See Hesterberg (2015) for an
  excellent overview of bootstrapping, a form of resampling with
  replacement that works well with sufficiently large sample sizes.} are
becoming increasingly popular because they provide a flexible approach
for a wide scope of problems, particularly those with small sample
sizes.

Unlike methods based on the Central Limit Theorem, permutation tests do
not generally require distributional (parametric) or sample size
conditions. Instead, they require \textbf{exchangeability}, an idea
which comprises much of the substance of our paper and which will be
introduced in Section~\ref{sec-exch}. Permutation tests were among the
first inferential tests conceived, and they were used widely (Fisher
1935) in the context of categorical data and two-sample comparisons.

Since their introduction, permutation tests have been extended to
address many different modeling contexts and are often presented to
support a comprehensive understanding of sampling distributions and
normal theory methods. As tools for testing that take full advantage of
the computer, permutation tests have become part of what we think of as
``computational statistics.'' The undergraduate statistics curriculum
has historically not had space for extended explication of permutation
methods; however, many modern introductory textbooks do introduce the
basics of permutation tests as part of the core content of the course
(Ismay and Kim 2020; Chance and Rossman 2021; Baumer et al. 2024;
Çetinkaya-Rundel and Hardin 2023; Lock et al. 2025). And while
permutation tests have been shown to reinforce understanding of sampling
distributions ({Garfield et al.} 2012; Taylor and Doehler 2015) and
other inferential concepts, the \emph{theory} that underlies
permutations tests is typically missing from the conversation. In our
experience, students at the undergraduate level are rarely given any
guidance on the technical conditions required for permutation tests. To
be fair, the presentation of permutation tests in undergraduate texts is
almost always given under the \emph{iid} (independent and identically
distributed) setting, which is sufficient for permutation tests
conditions. However, the \emph{iid} setting excludes more complex
scenarios for which permutation tests can be quite useful.

In what follows, we describe and detail the idea of exchangeability,
which is the core technical condition underlying permutation tests. We
argue that working through data scenarios that go beyond two-samples or
have a non-\emph{iid} structure provides a foundation of exchangeability
to further develop permutation tests and hypothesis testing in general.
Our goal is to present the key ideas underlying permutation methods,
explore additional complications, and motivate how and why an
investigation into permutation-based inference is valuable as a way to
expand understanding of inference and statistical foundations.

\subsection{Exchangeability}\label{sec-exch}

At the foundation of valid permutation tests is the condition of
\emph{exchangeability} (Pitman 1937; Good 2002). While there are
typically no distributional or sample size restrictions on a permutation
test, recognizing how to incorporate the exchangeability condition can
sometimes be difficult (Welch 1990).

We use the following definition of exchangeability throughout:

\begin{quote}
Data are \textbf{exchangeable under the null hypothesis} if the joint
distribution from which the data came is the same before a permutation
as after a permutation when the null hypothesis is true.
\end{quote}

Good (2002) describes how permutation inference holds under
``exchangeability-preserving transforms, asymptotic exchangeability,
partial exchangeability, and weak exchangeability.'' His work, however,
is quite theoretical and beyond the scope of a typical undergraduate
course. Draper et al. (1993) gives a detailed account of the issues of
exchangeability, including more complicated settings, such as serial
correlation. Instead of proving theoretical properties of different
exchangeability conditions, we walk the reader through examples that
highlight different forms of exchangeability and violations of
exchangeability.

\subsection{Motivating example}\label{motivating-example}

Consider an example whose goal is to model performance in college as a
quantitative outcome variable (\(Y\)) based on whether the student took
Advanced Placement (AP) courses in high school (\(X_1,\) the dichotomous
``treatment'' or ``group'' variable) and the student's family income
level (\(X_2,\) a potential confounder). We begin by first assessing the
association between AP course taking and college performance using data
from a sample. For the moment, we leave aside \(X_2\), and we assume
that the families are independent and that only one student is included
from each family. This example is then equivalent to a comparison of two
groups, students who took AP courses and those who did not.

\section{Permutation tests and
exchangeability}\label{permutation-tests-and-exchangeability}

\subsection{How to permute?}\label{how-to-permute}

Carrying out a permutation procedure involves calculation of a statistic
for each of the permuted samples. The observed statistic is then
compared to the permuted null distribution where the association between
\(Y\) and \(X_1\) has been broken.

Many settings, including a two-sample comparison, can be modeled using
linear regression. For linear regression models with a single predictor
(simple linear regression, SLR), equivalent permutation inference
results will ensue no matter whether the outcome variable (\(Y\)) or the
predictor variable (\(X_1\)) is permuted.

\begin{figure}[tbhp]

\centering{

\includegraphics[width=0.5\linewidth,height=\textheight,keepaspectratio]{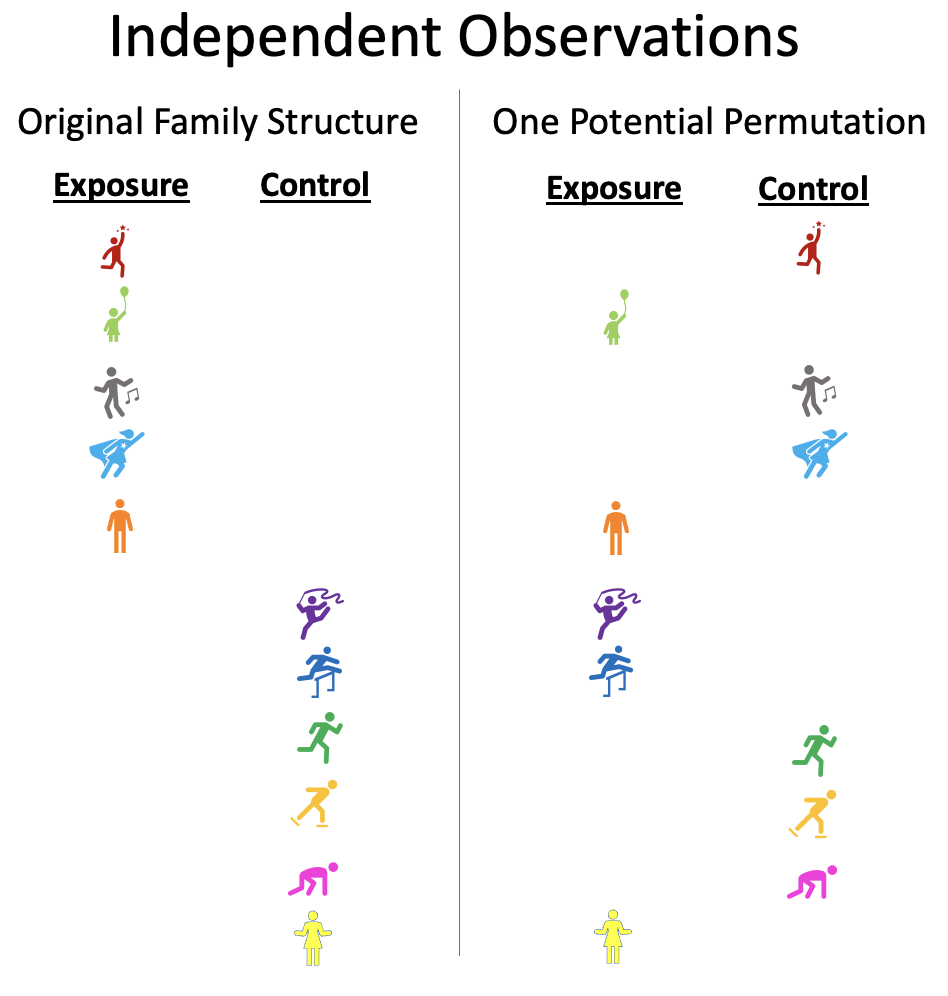}

}

\caption{\label{fig-indep-perm}Each color and row represents a family
structure. There are eleven individuals, none of whom are in the same
family. Note that in the permutation setting, the exposure group and the
control group both retain the same number of individuals as in the
original family structure.}

\end{figure}

Figure~\ref{fig-indep-perm} displays an example where interest is in
comparing the mean across two groups and the observations are
independent. There are eleven subjects, each from a different family;
five of the subjects are exposed (\(X_1=1,\) e.g., the exposure group
takes an AP course) and six of the subjects are control samples
(\(X_1=0\), e.g., no AP course). The outcome variable (\(Y,\) e.g.,
college performance) is measured for all subjects.

In the example, we can either permute the outcome \(Y\) or the group
indicator \(X_1\). We note that because we are permuting, there are
always five exposed observations and six control observations, leading
to \({11 \choose 6}\) = 462 possible permutations (although typically,
in this or larger sample sizes, we won't calculate all possible
permutations and instead sample from the distribution). We fit a SLR for
each permutation and use the resulting null distribution of the
t-statistics (i.e., the standardized slope coefficients for \(X_1\)) to
assess the significance of the treatment (grouping) variable.

\subsection{Clustered experimental design}\label{sec-cluster}

Understanding exchangeability is sometimes best understood when it is
violated. Consider a scenario where the underlying data are clustered in
some fashion. For example, consider a situation where, for convenience
or design purposes, families might be sampled rather than individuals.
So instead of having just one individual from each family, we have two
students from some families.

One of the conditions for inference in regression is that the
observations (or more accurately, the residuals) are independent of one
other. In the clustered data setting, it may no longer be tenable to
assume that individuals within a family are independent of others within
the same family.

Principled and flexible mathematical methods have been developed for a
variety of models in the setting where the observations are clustered or
repeated measures (see for example Laird and Ware (1982)). Random
intercept models, where each observational unit (e.g., family) is
assumed to have an underlying (unobserved) random level, are commonly
used to account for clustering. However, such mixed effects models are
both complicated (mixed effects models are not typically seen in
undergraduate curricula) and may require a sufficient sample size to
assume asymptotic normality of regression coefficients (Maas and Hox
2005).

Permutation methods are particularly desirable with clustered data since
they can be adapted to more complicated settings where exchangeability
might not be as straightforward as it is in the multiple linear
regression (MLR) setting (Draper et al. 1993). In the next subsection,
we detail how violations of exchangeability in the clustered data
setting can have a substantial impact on inferential results.

We revisit our motivating example where we are interested in predicting
the outcome (college performance, \(Y\)) as a function of a dichotomous
treatment (AP course taking, \(X_1\)) where some of the families have
two individuals within the sample. An adaptation of the permutation
tests that accounts for clustering within observational unit is a
computational alternative to the large-sample (i.e., based on the
Central Limit Theorem) random effects approach. The adapted permutation
test is attractive because it maintains an appropriate Type I error rate
(Winkler et al. 2014, 2015). Here we explore the clustered permutation
idea with two different data scenarios, representing two different types
of clustering that violate the independent observation condition.

\subsubsection{Cluster: permuting homogeneous clustered
observations}\label{cluster-permuting-homogeneous-clustered-observations}

\begin{figure}[tbhp]

\centering{

\includegraphics[width=0.5\linewidth,height=\textheight,keepaspectratio]{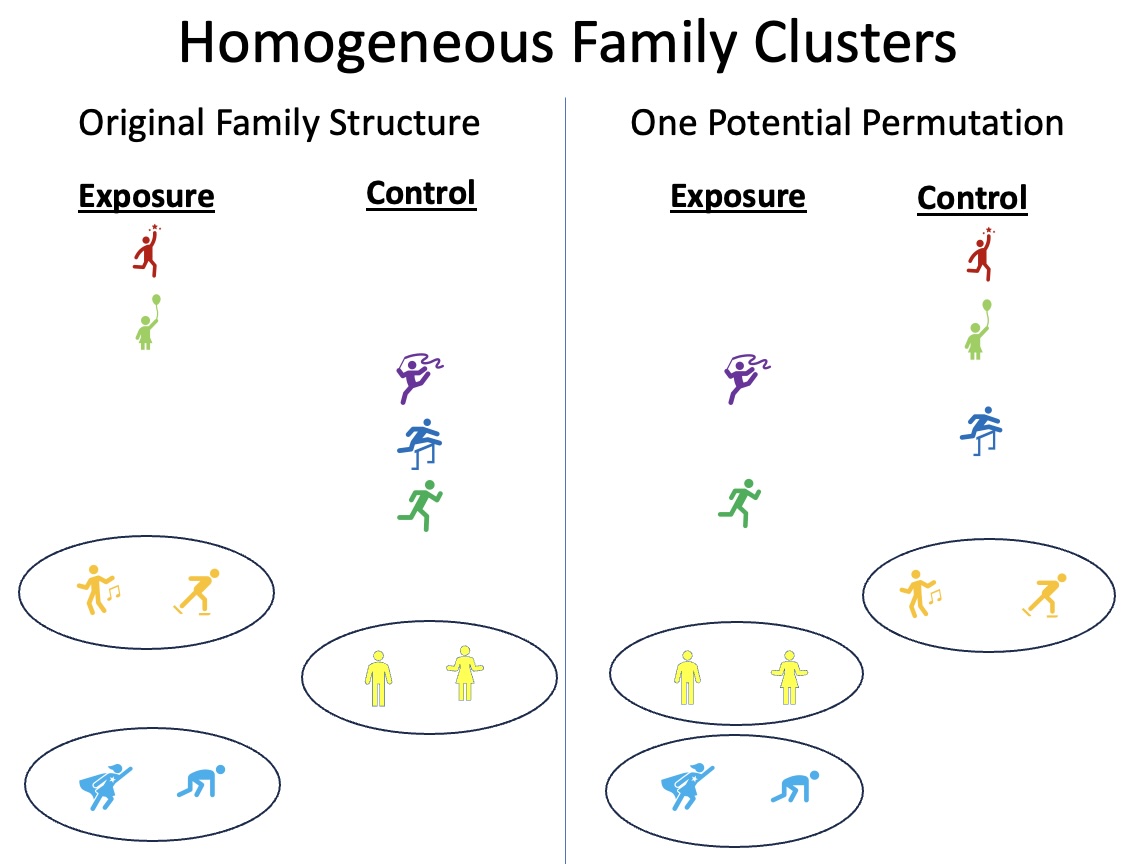}

}

\caption{\label{fig-homo-perm}Each color and row represents a family
structure. There are five single-person ``families'' and three families
with two members each. In the original structure, treatment happens by
family. Note that in the permutation setting, the exposure group and the
control group both retain the same number of individuals as in the
original family structure. The families are permuted together so as to
keep intact the original scenario of treatment within family.}

\end{figure}

In the homogeneous clustered (i.e., all individuals within a family have
the same value of the treatment variable) data scenario, there are
multiple observations of the same type within each observational unit
(see Figure~\ref{fig-homo-perm}). There are a total of eight families.
Five of the families have a single observation, while three of the
families have two individuals.

With homogeneous clustered observations, the permutation scheme must
preserve the family structure determined by the experimental design.
Here, each family has only one treatment (either exposure or control),
so we permute the treatments with the constraint that three of the
single-person families and one of the two-person families are in the
control group. For the homogeneous clustered example, there are
\({5 \choose 3} \cdot {3 \choose 1}\) = 30 possible permutations. We fit
a regression for each permutation and use the resulting null
distribution of the t-statistics (i.e., the standardized slope
coefficients for \(X_1\)) to assess the significance of the treatment
variable.

\subsubsection{Cluster: permuting heterogeneous clustered
observations}\label{cluster-permuting-heterogeneous-clustered-observations}

\begin{figure}[tbhp]

\centering{

\includegraphics[width=0.5\linewidth,height=\textheight,keepaspectratio]{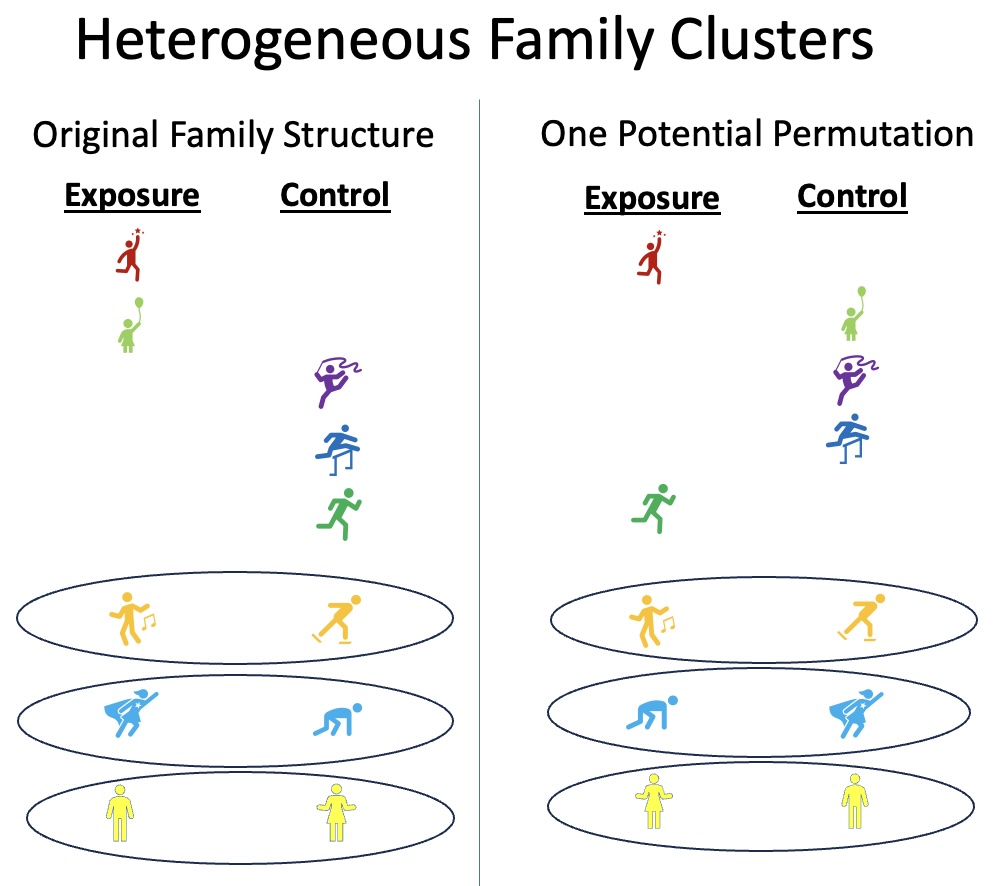}

}

\caption{\label{fig-hetero-perm}Each color and row represents a family
structure. There are five single-person ``families'' and three families
with two members each. In the original structure, for the families of
size two, exactly one member of each family is exposed, and the other
member of the family is in the control group. Note that in the
permutation setting, the exposure group and the control group both end
up with the same number of individuals as in the original family
structure. For the families of size two, the permutation happens within
the family so as to keep intact the original structure of both
treatments across the family.}

\end{figure}

In the heterogeneous clustered (e.g., each family has members with both
values of the dichotomous treatment) data scenario, families with two
observations have individuals of each type of treatment observed within
each observational unit (see Figure~\ref{fig-hetero-perm}). There are a
total of eight families. Five of the families have a single observation,
and three of the families have two individuals.

With heterogeneous clustered observations, the permutation scheme must
preserve the family structure. Here, the treatment on single-person
families will be permuted as if the observations were independent (i.e.,
a permutation with the constraint that two individuals are exposed and
three are control). The permutation of observations in the two-person
families happens within the family. That is, with equal probability, the
treatment levels are either kept the same or swapped, within each
family.

For the heterogeneous clustered scenario, there are
\({5 \choose 3} \cdot 2^3\) = 80 possible permutations. After each
permutation, we fit a regression on the quantitative predictor and the
permuted treatment. The resulting null distribution of the t-statistics
is used to assess the significance of the treatment variable.

\subsubsection{Simulation of clustered and independent
scenarios}\label{simulation-of-clustered-and-independent-scenarios}

With the clustered family structure, if the analyst permutes either the
group variable \(X_1\) or the outcome \(Y\) naïvely (without regard to
the family structure), there is a type of exchangeability violation due
to the independence (or lack thereof) of observations. The independence
exchangeability violation does matter (in terms of performance) as seen
in what follows.

We undertook a limited series of simulation studies to explore the
behavior of permutation tests when the design structure was
intentionally specified to violate the independence condition. The
simulations exploited the fact that ignoring clustering for the
homogeneous clusters yields standard errors that are too small and that
doing so for the heterogeneous cluster design, where family members
serve as their own control, yields standard errors that are too large
(Cannon et al. 2001).

The heart of our simulations is well described by Winkler et al. (2015),
who focus on

\begin{quote}
any dataset with known dependence among observations. In such cases,
some permutations, if performed, would create data that would not
possess the original dependence structure, and thus, should not be used
to construct the reference (null) distribution. To allow permutation
inference in such cases, we test the null hypothesis using only a subset
of all otherwise possible permutations, i.e., using only the
rearrangements of the data that respect exchangeability, thus retaining
the original joint distribution unaltered.
\end{quote}

For each model/scenario combination, 2000 simulations were generated
with \(n = 20\) subjects per treatment group (more subjects than
displayed in Figure~\ref{fig-indep-perm}, Figure~\ref{fig-homo-perm},
and Figure~\ref{fig-hetero-perm}). For the clustering scenarios, half of
the subjects were singletons and the others shared a family member
within the group (homogeneous) or across groups (heterogeneous). Within
each simulation, the null distribution was created using 2000
permutations.

Table~\ref{tbl-simresults} displays the results from the simulation
study. The \texttt{binom.test()} function in R was used to calculate a
95\% CI for the estimated Type I error rates. As expected, the models
that do not appropriately account for clustering are anti-conservative
(``lm homogeneous'', ``naïve permutation homogeneous'') or overly
conservative (``lm heterogeneous'' and ``naïve permutation
heterogeneous''); we have bolded the models that don't maintain the
desired Type I error rate of 0.05. Methods that appropriately account
for the clustering (e.g., ``lme'' or ``correct permutation'') include
the desired Type I error rate within the 95\% CI. When there are cluster
dependencies, the naïve permutation method is not advised.

\begin{longtable}[t]{lll}

\caption{\label{tbl-simresults}Type I error rates for different
scenarios (2000 simulations, each with 2000 permutations for each
simulation). Models include SLR (lm), linear mixed effects (lme),
permutation not accounting for clustering (naïve), and permutation
accounting for clustering (correct). Forms of clustering are independent
(yellow), homogeneous (white), and heterogeneous (blue). The dichotomous
treatment variable was permuted (Draper and Stoneman 1966). Type I error
was computed using \(\alpha = 0.05.\)}

\tabularnewline

\toprule
Model/Scenario & Type I error rate (95\% CI) & Results\\
\midrule
\cellcolor[HTML]{FBF5DF}{lm independent} & \cellcolor[HTML]{FBF5DF}{0.042 (0.034-0.052)} & \cellcolor[HTML]{FBF5DF}{includes desired alpha level}\\
\cellcolor[HTML]{FBF5DF}{correct permutation independent} & \cellcolor[HTML]{FBF5DF}{0.046 (0.037-0.056)} & \cellcolor[HTML]{FBF5DF}{includes desired alpha level}\\
\textbf{lm homogeneous} & \textbf{0.07 (0.059-0.082)} & \textbf{anticonservative}\\
lme homogeneous & 0.051 (0.041-0.061) & includes desired alpha level\\
\textbf{naïve permutation homogeneous} & \textbf{0.072 (0.061-0.084)} & \textbf{anticonservative}\\
\addlinespace
correct permutation homogeneous & 0.045 (0.036-0.055) & includes desired alpha level\\
\textbf{\cellcolor[HTML]{F5FEFD}{lm heterogeneous}} & \textbf{\cellcolor[HTML]{F5FEFD}{0.029 (0.023-0.038)}} & \textbf{\cellcolor[HTML]{F5FEFD}{conservative}}\\
\cellcolor[HTML]{F5FEFD}{lme heterogeneous} & \cellcolor[HTML]{F5FEFD}{0.044 (0.035-0.054)} & \cellcolor[HTML]{F5FEFD}{includes desired alpha level}\\
\textbf{\cellcolor[HTML]{F5FEFD}{naïve permutation heterogeneous}} & \textbf{\cellcolor[HTML]{F5FEFD}{0.034 (0.027-0.043)}} & \textbf{\cellcolor[HTML]{F5FEFD}{conservative}}\\
\cellcolor[HTML]{F5FEFD}{correct permutation heterogeneous} & \cellcolor[HTML]{F5FEFD}{0.056 (0.047-0.068)} & \cellcolor[HTML]{F5FEFD}{includes desired alpha level}\\
\bottomrule

\end{longtable}

\subsection{Multiple linear
regression}\label{multiple-linear-regression}

It is well documented that work in AP courses is positively associated
with socio-economic status (Kolluri 2018), which motivates the
introduction of \(X_2\) as a possible confounder of the relationship
between \(X_1\) and \(Y\) (Cummiskey et al. 2020). Throughout all of the
four MLR hypothesis tests described below, assume that the null
hypothesis (\(\beta_{1} = 0\)) is true, that is, having taken an AP
course (\(X_1\)) is not linearly related to performance in college
(\(Y\)) after controlling for family income (\(X_2\)).

Unlike SLR, with MLR, there is no obvious choice of how to permute. In
what follows, we discuss four different permutation methods and their
exchangeability conditions. The goal of permuting is to create a null
sampling distribution of the statistic of interest so that we may infer
the variability of the statistic. That is, a distribution of the
statistics under the setting where the outcome \(Y\) and the predictor
of interest \(X_1\) are not linearly related. But the null distribution
also needs to be created under the exchangeability condition---the
permuting only leads to a change in the relationship between \(Y\) and
\(X_1\) without changing any of the other variable relationships in the
linear model.

Communicating the different permutation structures to students is an
ideal way to explore the ideas of exchangeability. Even though the
methods below are quite similar with respect to Type I errors and power,
the \emph{discussion} of the methods allows students to study
exchangeability and permutations more generally. We include extended
details of the permutation schemes in Appendix~\ref{sec-appendix-perm}.
Here, we briefly describe the four permutation methods with specific
thought to the exchangeability violations for each one.

\subsubsection{\texorpdfstring{Permute
\(\mathbf{Y}\)}{Permute \textbackslash mathbf\{Y\}}}\label{permute-mathbfy}

At first glance, it might seem like permuting the outcome variable would
be a good way to break the relationship between \(Y\) and \(X_1\) (Manly
1986, 1997) (see complete algorithm in Appendix
\ref{sec-appendix-perm}). Indeed, permuting \(Y\) will break the
relationship between \(Y\) and \(X_1,\) which will force the null
hypothesis to be true (which is what we want for testing). Additionally,
permuting \(Y\) will preserve the relationship between \(X_1\) and
\(X_2\) (as desired). However, permuting \(Y\) will also simultaneously
break the relationship between \(Y\) and \(X_2,\) which may not be
acceptable if we need to preserve the relationship to mirror the
original data structure.

Breaking the relationship between \(Y\) and \(X_2\) wouldn't be an issue
if \(Y\) and \(X_2\) were not associated. But if \(X_2\) is a potential
confounder of the relationship, then \(Y\) and \(X_2\) may be
associated. The permutation procedure where the outcome \(Y\) is
permuted, therefore, is not exchangeable, even when the null hypothesis
of interest is true. In this setting, permuting the outcome creates a
permutation sampling distribution with the unintended consequence that
performance in college (\(Y\)) is no longer associated with income level
(\(X_2\)). Because performance and income \textbf{are} associated in the
real world, the data with permuted \(Y\) values do not retain the
observed dependency structure present in the original data. That is, a
relevant relationship is lost after a permutation, which ends up
violating the exchangeability condition.

The first block of rows in table Table~\ref{tbl-perms} summarizes the
broken and preserved relationships when permuting \(Y.\) Note that when
\(Y\) is permuted, the original relationship between \(X_1\) and \(X_2\)
is preserved (which is what we want in terms of exchangeability). The
broken relationship between \(X_2\) and \(Y,\) however, is problematic
in terms of exchangeability. If we permute \(Y,\) then exchangeability
is met \textbf{only} if \(Y\) and \(X_2\) are uncorrelated in the
original dataset.

\begin{longtable}[]{@{}
  >{\raggedright\arraybackslash}p{(\linewidth - 6\tabcolsep) * \real{0.3363}}
  >{\raggedright\arraybackslash}p{(\linewidth - 6\tabcolsep) * \real{0.2389}}
  >{\raggedright\arraybackslash}p{(\linewidth - 6\tabcolsep) * \real{0.2124}}
  >{\raggedright\arraybackslash}p{(\linewidth - 6\tabcolsep) * \real{0.2124}}@{}}
\caption{Multiple linear regression models with two covariates; we show
different permutation schemes, variable relationships that are broken,
and variable relationships that are preserved. Any relationships that
are broken, and \textbf{not} null, will violate exchangeability. Note
that all methods require the observations to be independent of one
another. See Appendix \ref{sec-appendix} for notation and full
permutation schemes.}\label{tbl-perms}\tabularnewline
\toprule\noalign{}
\begin{minipage}[b]{\linewidth}\raggedright
Permutation
\end{minipage} & \begin{minipage}[b]{\linewidth}\raggedright
Broken Relationships
\end{minipage} & \begin{minipage}[b]{\linewidth}\raggedright
Preserved Relationships
\end{minipage} & \begin{minipage}[b]{\linewidth}\raggedright
Permutation distribution
\end{minipage} \\
\midrule\noalign{}
\endfirsthead
\toprule\noalign{}
\begin{minipage}[b]{\linewidth}\raggedright
Permutation
\end{minipage} & \begin{minipage}[b]{\linewidth}\raggedright
Broken Relationships
\end{minipage} & \begin{minipage}[b]{\linewidth}\raggedright
Preserved Relationships
\end{minipage} & \begin{minipage}[b]{\linewidth}\raggedright
Permutation distribution
\end{minipage} \\
\midrule\noalign{}
\endhead
\bottomrule\noalign{}
\endlastfoot
\textbf{Permute \(Y\)} & \(X_1\) \& \(Y\) & \(X_1\) \& \(X_2\) &
\(t^* = \frac{b_{1\cdot2}^* - 0}{SE(b_{1\cdot2}^*)}\) \\
Manly (1986, 1997) & \(X_2\) \& \(Y\) & & See Eq (\ref{t_y}) \\
-------------------------------------- & --------------------------- &
------------------------ & ------------------------ \\
\textbf{Permute \(X_1\)} & \(X_1\) \& \(X_2\) & \(X_2\) \& \(Y\) &
\(t^* = \frac{b_{1\cdot2}^* - 0}{SE(b_{1\cdot2}^*)}\) \\
Draper and Stoneman (1966) & \(X_1\) \& \(Y\) & & See Eq (\ref{t_x1}) \\
-------------------------------------- & --------------------------- &
------------------------ & ------------------------ \\
\textbf{Permute reduced model residuals} & \(X_1\) \& \(Y\) (if \(X_1\)
\& \(X_2\) are uncorrelated) & \(X_1\) \& \(X_2\) &
\(t^* = \frac{b_{1\cdot2}^* - 0}{SE(b_{1\cdot2}^*)}\) \\
Freedman and Lane (1983) & & \(X_2\) \& \(Y\) & See Eq (\ref{t_red}) \\
-------------------------------------- & --------------------------- &
------------------------ & ------------------------ \\
\textbf{Permute full model residuals} & None & \(X_1\) \& \(X_2\) &
\(t^* = \frac{b_{1\cdot2}^* - b_{1\cdot2}}{SE(b_{1\cdot2}^*)}\) \\
ter Braak (1990, 1992) & & \(X_1\) \& \(Y\) & See Eq (\ref{t_full}) \\
& & \(X_2\) \& \(Y\) & \\
\end{longtable}

\subsubsection{\texorpdfstring{Permute
\(\mathbf{X_1}\)}{Permute \textbackslash mathbf\{X\_1\}}}\label{permute-mathbfx_1}

In order to maintain the relationship between \(Y\) and \(X_2\) (while
still interested in the relationship between \(Y\) and \(X_1\)), we
might consider permuting \(X_1\) instead of \(Y\) (Draper and Stoneman
1966) (see complete algorithm in Appendix \ref{sec-appendix-perm}).
Indeed, the permutation distribution created from permuting \(X_1\) will
force the null hypothesis to be true. However, permuting \(X_1\) has the
side effect of breaking the relationship between \(X_1\) and \(X_2\) in
the permuted data. If the data come from, for example, a randomized
clinical trial (where \(X_1\) is the treatment variable), then \(X_1\)
and \(X_2\) will be independent in the original dataset, and permuting
\(X_1\) will on average \textbf{not} violate the exchangeability
condition. If \(X_1\) and \(X_2\) are correlated in the original
dataset, as seen in our AP course and socio-economic status example,
then permuting \(X_1\) violates the exchangeability condition. The
second block of tows in Table~\ref{tbl-perms} summarizes the broken and
preserved relationships when permuting \(X_1.\) Note that when \(X_1\)
is permuted, the original relationship between \(X_2\) and \(Y\) is
preserved (which is what we want in terms of exchangeability). The
broken relationship between \(X_1\) and \(X_2,\) however, is problematic
under general conditions.

\subsubsection{Permute reduced model
residuals}\label{sec-perm-red-resid}

In order to address some of the concerns of the first two MLR
permutation methods, Freedman and Lane (1983) propose a permutation
scheme based on residuals (see complete algorithm in Appendix
\ref{sec-appendix-perm}). The permutation preserves the relationship
between \(X_1\) \& \(X_2\) as well as the relationship between \(X_2\)
\& \(Y.\) However, in order for the relationship between \(X_1\) \&
\(Y\) (the key relationship!) to be broken, \(X_1\) and \(X_2\) must not
be associated. The third block of rows in Table~\ref{tbl-perms}
summarizes the broken and preserved relationships when permuting the
reduced model residuals. Note that when permuting the reduced model
residuals, the original relationships between both \(X_1\) \& \(X_2\)
and additionally between \(X_2\) \& \(Y\) are preserved (which is what
we want in terms of exchangeability). The null hypothesis is true only
if the relationship between \(X_1\) and \(Y\) is broken, and that
happens only when \(X_1\) and \(X_2\) are uncorrelated. Appendix
\ref{sec-appendix-proof} sketches the dependence of \(\rho(Y^*, X_1)\)
(the correlation between the permuted \(Y\), \(Y^*\), and \(X_1\)) on
\(\rho(X_1, X_2)\) (the correlation between \(X_1\) and \(X_2).\)

\subsubsection{Permute full model residuals}\label{sec-perm-full-resid}

As an extension to Freedman and Lane (1983), ter Braak's (1990, 1992)
permutation method permutes the residuals from the full model (see
complete algorithm in Appendix \ref{sec-appendix-perm}). The last block
of rows in Table~\ref{tbl-perms} summarizes the broken and preserved
relationships when permuting the full model residuals. Note that
permuting the residuals under the full model allows all of the
exchangeability conditions to hold. Unlike the previous three
permutation schemes, here, the new model does not force the null
hypothesis to be true, which is why the test statistic measures the
deviation of the coefficients (on \(X_1\)) from the permuted data to the
original data model coefficient (on \(X_1\)), instead of comparing to
zero. The procedure works because the variability of the permuted
coefficient around the sample coefficient mimics the variability of the
sample coefficient around the parameter ({ter Braak} 1990). See Appendix
\ref{sec-alg-full} for a more full justification of the use of the
permutation statistic.

\subsubsection{MLR: does it matter?}\label{mlr-does-it-matter}

What impact does violation of exchangeability have on MLR inferences? We
might consider studies of Type I error rate (when the null hypothesis is
true) as well as Type II error rate (when the alternative hypothesis is
true). Prior research (Anderson and Legendre 1999; Winkler et al. 2014)
has shown that hypothesis testing in MLR is quite robust to the choice
of permutation, and, except in cases of extreme error distributions,
meeting (or not) the exchangeability conditions (such as breaking the
association between \(Y\) and \(X_2\)) does not substantially impact the
size or power of the test.

\subsection{Comparison of correlation}\label{comparison-of-correlation}

A fourth example to help students think more deeply about
exchangeability comes from a setting that does not have a parallel
commonly taught standard mathematical test\footnote{The test of equality
  of correlations in two groups can be carried out using Fisher's
  \(r\)-to-\(z\) transformation (Fisher 1921; Diedenhofen and Musch
  2015) but only with large samples (Stanley et al. 2024), and the
  procedure is not typically taught in an undergraduate curriculum.}.
Consider the case where interest is in comparing the population
correlation (\(\rho\)) across groups A (\(X_1 = 1\)) and B
(\(X_1 = 0\)).

\[H_0: \rho_A = \rho_B\] \[H_A: \rho_A \ne \rho_B\] The dataset includes
three columns: one quantitative variable (\(X_2\)), the other
quantitative variable (\(Y\)), and the group variable (\(X_1\): Group A
or Group B). A first glance at permuting would suggest to permute the
group variable and use \(r^*_A - r^*_B\) (the difference in sample
correlations computed after permutation) as the test statistic given in
the null sampling distribution. However, the permutation scheme violates
exchangeability \emph{unless} the models in \(A\) and \(B\) are the same
in every way except the correlation.

\begin{figure}[tbhp]

\centering{

\includegraphics[width=1.1\linewidth,height=\textheight,keepaspectratio]{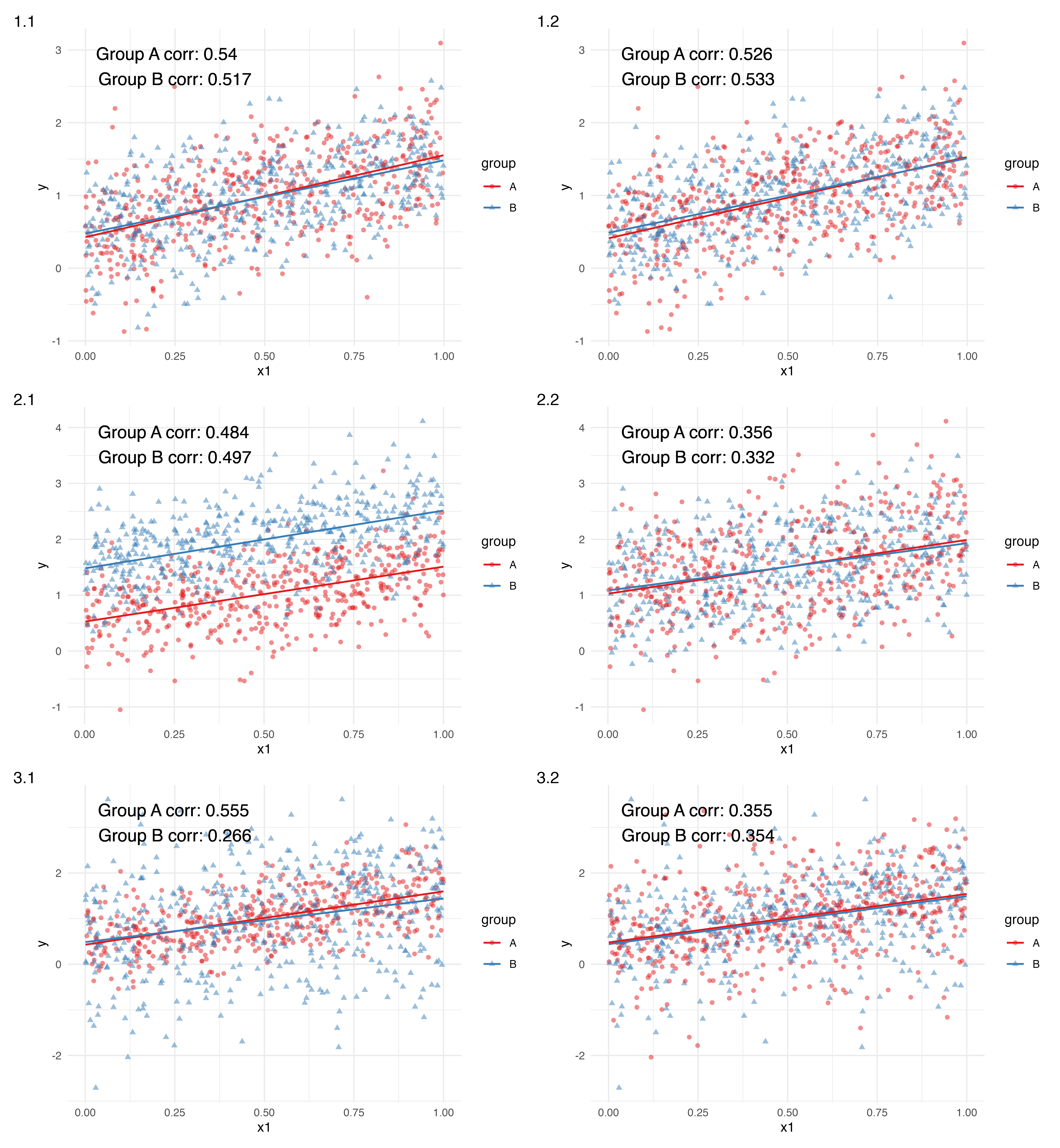}

}

\caption{\label{fig-corr}Results from 1. a setting where exchangeability
holds and correlations are the same (Figure 4.1.1 displays sample data
and Figure 4.1.2 displays a permuted sample); 2. a setting where the
null hypothesis is true but exchangeability does not hold because the
intercepts are different (Figue 4.2.1 displays sample data and Figure
4.2.2 displays a permuted sample with attenuated difference in
correlations); and 3. a setting where exchangeability holds but
correlations are different (4.3.1 displays sample data and 4.3.2
displays a permuted sample with attenuated difference in correlations).}

\end{figure}%

Figure~\ref{fig-corr} illustrates ideas of both exchangeability and the
null hypothesis. Figure 4.1.1 displays a sample from a world where the
intercept, slope, and correlation are the same in groups A and B (i.e.,
the null hypothesis is true and all exchangeability conditions are met).
Figure 4.1.2 displays the data and correlations in permuted groups.
Because there is no difference between groups and exchangeability holds,
the results from Figures 4.1.1 and 4.1.2 are quite similar.

However, Figures 4.2.1 and 4.2.2 highlight an arguably more common
situation where the linear model across the two groups differs. In
Figure 4.2.1, the null hypothesis happens to be true (the correlations
are the same), but the violation of exchangeability (different
intercepts) is important to notice regardless of whether or not the null
hypothesis is true. Recall the definition of exchangeability which
states that the data are exchangeable under the null hypothesis only if
their distribution is the same before and after permutation. In Figure
4.2.1, the null hypothesis is true, and the distribution of the permuted
data (Figure 4.2.2) are clearly not the same as the original data. If
exchangeability is violated, then the process of permuting the group
variable may not lead to a valid hypothesis test (regardless of whether
or not the null hypothesis is true).

Contrast this with Figures 4.3.1 and 4.3.2, which represent data where
the intercept and slope are the same but the correlations are different
(i.e., the null hypothesis is not true but exchangeability holds). The
estimates of the difference in correlations from the permutation in
Figure 4.3.2 is smaller than the observed sample in Figure 4.3.1.
Permuting the group variable would not violate exchangeability because
the models are identical except for the difference in correlation (i.e.,
except for the condition of the null hypothesis).

We carried out a simulation study to assess the impact of violations of
the exchangeability assumption in this setting. We ran 5,000 simulations
with n = 20 observations in each of the two groups. The true regression
model was defined as \(\beta_0\) = 0.5, \(\beta_1 = 1\) with root MSE =
0.5. Here, the underlying correlation is 0.5 in both groups
(corresponding to Figures 4.1.1 and 4.1.2). The permutation test which
shuffled the group indicator and calculated the difference in
correlations effectively maintained the desired Type 1 error rate of
0.05 (average Type I error of 0.052, 95\% CI ranged from 0.046 to
0.059).

We next considered the situation described in the second row of
Figure~\ref{fig-corr} (Figures 4.2.1 and 4.2.2), where the correlations
remain the same (0.5 in both groups), but the intercepts are different
(0.5 in group A and 1.5 in group B). In this scenario, where
exchangeability is violated, the permutation procedure fails to maintain
an appropriate Type I error rate (average Type I error of 0.024, 95\% CI
ranged from 0.020 to 0.029). Notice that the correlations in Figure
4.2.2 are substantially smaller than the correlations in Figure 4.2.1
(as expected). The variability of the correlation is inversely related
to the magnitude of the correlation. Which is to say that the permuted
null sampling distribution (from permutations like those seen in Figure
4.2.2) will be wider than the theoretical null sampling distribution
from data like those in Figure 4.2.1. Because the permuted null sampling
distribution is wider, the Type I error rate is lower than the nominal
0.05 level.

For the bottom row of Figure~\ref{fig-corr} (Figures 4.3.1 and 4.3.2),
we did not test Type I error because the null hypothesis is false. We
could have done a power analysis, but assessing power under various
settings was beyond the scope of the paper.

There are multiple other linear regression settings where students can
consider exchangeability violations and hypothesis conditions. For
example, an extension of Figure~\ref{fig-corr} would be to create a
setting where the slopes differed and the alternative hypothesis was
true (e.g., residuals are bigger in one plot). Additionally, changing
the range of the \(X_1\) values across Group A and Group B would also
violate exchangeability (as would changing the range of the \(Y\) values
across Groups A and B).

\section{Discussion and pedagogical
recommendations}\label{discussion-and-pedagogical-recommendations}

Sometimes the easiest way to understand an assumption is to explore when
the assumption is violated. In this paper we provide four classes of
examples and discuss what exchangeability means in each. We discuss how
the analysis does or does not violate exchangeability, depending on the
permutation scheme itself. We hope that the examples prompt students to
ask themselves ``is this even valid given the question of interest and
properties of the data?'' when running a permutation test.

We do not provide the examples to suggest that permutation tests are or
are not superior to their mathematical counterparts. Indeed, as seen in
the MLR setting, the literature is clear that some violations of
exchangeability do not have a substantial impact on the Type I error.

Instead, we provide the examples in order to continue the conversation
surrounding permutation tests at the undergraduate level. We hope that
our examples will be used in the classroom as a way for teachers to
explain to students that permutation tests \textbf{do} have technical
conditions. Just like other tests in the introductory canon (e.g., the
t-test), the permutation test is more robust to violations of some
conditions and less robust to violations of others.

The sophisticated and extremely powerful concept of exchangeability can
be accessed through examples presented in the classroom. After digging
into the exchangeability ideas, students are able to apply a permutation
approach to more complicated data settings (which are often encountered
in real applications). Our approach is consistent with the advice of
Chance et al. (2024), who hoped ``to help instructors understand the
differences in the simulation strategies to better inform their own
decisions of how to adapt a simulation approach for their classes and to
better respond to student questions that may arise.''

Another compelling rationale (from Vanhove (2015)) to devote time to
these questions is the concept that ``the test follows from the
design'':

\begin{quote}
There is a clear logical link between the statistical test we use and
the experimental design we opted for: using a permutation test is
entirely warranted by the random assignment of participants to two
equal-sized groups. Stressing the link between experimental design and
statistical inference -- rather than considering them separately -- is
of huge pedagogical, as well as practical, use, I believe (Vanhove
2015).
\end{quote}

In randomized studies, the link between experimental design and
permutation scheme should be made clear to students. Shuffling treatment
labels models random assignment of a single variable, while the
clustered schemes model restricted randomization. An understanding of
the experimental design can bring the appropriate permutation scheme to
light.

\subsection{Classroom recommendations}\label{classroom-recommendations}

\paragraph{Experience teaching
exchangeability}\label{experience-teaching-exchangeability}

In a recent upper division undergraduate statistics elective class, one
of the authors presented some of the ideas described in the paper. In
particular, she asked her students how permuting \(X_1\) violates
exchangeability in the MLR setting. The answer starts by acknowledging
that when \(X_1\) is permuted, the relationship between \(Y\) and
\(X_1\) is broken, which is necessary for \(H_0\) to be true. However,
the relationship between \(X_1\) and \(X_2\) is also broken, and if
\(X_1\) and \(X_2\) were correlated in the original data, they should be
correlated in the permuted data as well (so exchangeability is broken).
The following represent paraphrases of the type of answers students gave
to the in-class activity. The responses are illustrative anecdotes; they
are not meant to indicate a proportional representation of the suite of
potential answers. Student-paraphrased responses are given in quotes,
while our reflections are given in italics. The prompt given to the
students is:

\begin{quote}
Your friend proposes to permute the variable \(X_1\) in order to create
a null sampling distribution for the coefficient on \(X_1\). In what way
does your friend's permutation scheme violate the exchangeability
condition?
\end{quote}

\begin{itemize}
\tightlist
\item
  ``Exchangeability means that the data before and after the permutation
  comes from the same distribution\ldots{} we are incorrectly assuming
  that \(X_2 = 0,\) i.e., that \(X_2\) does not matter.'' \emph{The
  student is able to state the definition of exchangeability, but they
  confuse the idea of the coefficient being zero (\(H_0: \beta_2 = 0\))
  and the variable being zero. The student also misses the important
  aspect of the relationship between \(X_1\) and \(X_2.\)}
\item
  ``If we originally had correlation, we no longer have correlation.''
  \emph{The student understood that aspects of the relationships needed
  to be the same originally and in the permuted data, but they did not
  specify which variables should be correlated.}
\item
  ``Ignoring the other covariates, breaking the conditional structure
  for the test.'' \emph{The student conflates the conditional structure
  of the data itself with some aspect of the test.}
\item
  ``If \(X_1\) and \(X_2\) are correlated then permuting \(X_1\) will
  break that correlation / relationship.'' \emph{The student understands
  the issue of violating exchangeability but does not discuss the
  motivation of permuting \(X_1\) to create the null sampling
  distribution.}
\item
  ``By permuting \(X_1,\) we break the \(Y\) and \(X_1\) relationship,
  but now we have also broken the relationship between \(X_1\) and
  \(X_2.\)'' \emph{The student is able to succinctly identify the
  problem.}
\end{itemize}

We close with some practical guidance about how to bring exchangeability
and permutation tests into the classroom. Underlying each of the
classroom activities is the learning goal of students assessing whether
the resampling scheme actually matches the question of interest. The
suggestions are organized by type of classroom.

\paragraph{introductory undergraduate}\label{introductory-undergraduate}

\begin{itemize}
\tightlist
\item
  Teacher creates a simulation program (Rmd/qmd file or Shiny App) where
  students can change the error structure within a MLR analysis and
  discover that the desired Type I error rates are not achieved only in
  the case when the errors are particularly egregious.
\item
  Use existing applets\footnote{e.g., the Two-variable ANOVA applet by
    Rossman/Chance
    \url{https://www.rossmanchance.com/applets/2021/twoway/twoway.html}}
  to walk through permuting different variables or residuals for
  inferential analyses.
\item
  Use a coin for an in-class permutation test activity on paired
  observations (e.g., see the ``Matched Pairs Applet'' described in
  Section 22.2 of Chance and Rossman (2021), with the corresponding URL
  of
  \url{https://www.rossmanchance.com/applets/2021/matchedpairs/MatchedPairs.htm}).
\end{itemize}

\paragraph{intermediate / advanced
undergraduate}\label{intermediate-advanced-undergraduate}

\begin{itemize}
\tightlist
\item
  Teach permutation tests for MLR. Talk to your students about the
  \textbf{choice} they will have to make (permute \(Y\)? \(X_1\)?
  residuals?) and that all modeling contains \textbf{choices}. Choices
  are not usually objective or unbiased, so an ability to defend their
  choice is what gives them power as a statistician.
\item
  Describe the difference between creating a null sampling distribution
  (i.e., ensuring the data distribution is the same before and after
  permuting) and establishing that exchangeability holds (i.e., making
  sure that the Type I error rates will be accurate). Have students
  describe which aspect of the permutation does which job.
\item
  Come up with scenarios (or use our cluster scenarios!) where the
  students can figure out the correct permutation schemes. See an
  example from JH's class that uses permutations to address a research
  question using a stratified two-sample test.\footnote{\url{https://st47s.com/Math154/Notes/permschp.html\#macnell-teaching-evaluations-stratified-two-sample-t-test}}
  Or have the students explore how to develop a test for symmetry of a
  distribution (e.g., the null hypothesis is that the distribution is
  symmetric).
\item
  Carry out a variety of permutation tests using a MLR model from an
  example dataset. Have students check the implementations or provide
  some of the implementations and have them carry out the others.
\end{itemize}

\paragraph{advanced undergraduate}\label{advanced-undergraduate}

\begin{itemize}
\tightlist
\item
  Ask your students what it means that ``the test follows from the
  design.'' Use the cluster scenarios and apply the opposite permutation
  scheme. Have them compare the permutation scheme and the experimental
  design to figure out what makes the most \textbf{sense} in terms of
  the relationship of the permuting and the design.
\item
  Ask students to read the paper by Helwig (2019) and explore the
  implications for the MLR setting via the \texttt{nptest} package in R
  (Helwig 2023). Have them explore how and why robust permutation tests
  (DiCiccio and Romano 2017) use alternative test statistics.
\item
  Have students run a simulation study to see if the Type I error rates
  given by the exchangeability violation in Figure~\ref{fig-corr} vary
  substantially from the level of significance.
\end{itemize}

\paragraph{advanced undergraduate / beginning
graduate}\label{advanced-undergraduate-beginning-graduate}

\begin{itemize}
\tightlist
\item
  Present some or all of the regression methods described in Appendix
  \ref{sec-appendix} (with more in Winkler et al. (2014)) and have
  students fill out a blank Table~\ref{tbl-perms} (see Appendix
  \ref{sec-appendix}). Ask students to report which relationships are
  broken and which are preserved.
\item
  Find permutation tests in the literature and have students describe
  how the permutation scheme upholds (or doesn't!) the exchangeability
  condition.
\end{itemize}

\paragraph{graduate}\label{graduate}

\begin{itemize}
\tightlist
\item
  Find theoretical work that proves exchangeability and demonstrate the
  important mapping of the theory to the applied problems (or maybe the
  exchangeability is not vital to the performance of the method?).
\end{itemize}

We suggest some ideas at particular levels, but many of the pedagogical
ideas can be adjusted to be effective at a variety of levels. Whether to
permute \(Y\) or \(X_1\) can be taught as early as introductory
statistics, where students have done some hypothesis testing and some
multivariate modeling. Graduate students can think carefully about how
to prove that conditions of exchangeability are met. The group of
students who might most benefit from the ideas we've presented include
upper level undergraduates and early graduate students who are focused
on understanding which models are best in which settings and how to
differentiate those settings.

\section{Acknowledgements}\label{acknowledgements}

We thank the reviewers and associate editor for many helpful suggestions
that have improved the manuscript.

\section{Disclosure statement}\label{disclosure-statement}

The authors report there are no competing interests to declare.

\section{Reproducibility statement}\label{reproducibility-statement}

The code used to undertake the simulation studies reported in the paper
have been made available at \url{https://osf.io/2avw6}.

\section*{References}\label{references}
\addcontentsline{toc}{section}{References}

\protect\phantomsection\label{refs}

\newpage

\appendix

\section{Appendix}\label{sec-appendix}

Here we review four different approaches for carrying out permutation
tests for MLR models. We provide some theoretical content that might be
of interest to an advanced undergraduate or early graduate student.

\subsection{Why permutation tests?}\label{why-permutation-tests}

As previously observed, even when both the normality conditions and the
exchangeability are modestly violated, all four of the permutation test
methods described below do reasonably well at controlling Type I errors
(Anderson and Legendre 1999; Winkler et al. 2014). When error terms come
from extremely skewed distributions (e.g., cubed exponentials), the
permutation tests maintained the appropriate Type I error rate while
normal theory methods were overly conservative (Anderson and Legendre
1999). There is strong evidence that in order to violate alpha-level
testing, the data must violate the technical conditions to an extreme
degree (Anderson and Legendre 1999; Tantawanich 2006; Winkler et al.
2014).

If we anticipate that results will be indistinguishable, why is an
understanding of permutation methods (and exchangeability) important?
Why do we present the methods below as valuable for communicating
exchangeability to our students?

\begin{enumerate}
\def\labelenumi{\arabic{enumi}.}
\tightlist
\item
  We believe that exchangeability is a valuable concept for students to
  understand, as it undergirds key foundational knowledge of statistical
  inference.
\item
  There do exist settings where mathematical (i.e., based on the Central
  Limit Theorem) and permutation approaches produce different
  inferential results, even within the linear model framework. Knowledge
  of permutation-based approaches (plus computational skills) allows
  inference in such areas where linear model conditions don't hold, and
  permutation tests may be straightforward to implement in situations
  where typical parametric tests may exist but be extremely complicated
  (e.g., dependence in the experimental design, as described in
  Section~\ref{sec-cluster}). As long as permuted draws can be made from
  the appropriate null world, permutation tests can be used to make
  inferences without parametric conditions.
\end{enumerate}

\subsection{Set-up}\label{set-up}

In order to work through the details of the permutation methods and
corresponding exchangeability constraints, we provide notation and model
specifications to describe the MLR model.

The population model of interest is
\[E[Y|X_1, X_2] = \beta_{0\cdot1,2} + \beta_{1\cdot2} X_1 + \beta_{2\cdot1} X_2,\]
where \(X_1\) is the variable of interest in predicting \(Y,\) and
\(X_2\) is a nuisance variable. The \(\cdot\) (dot) notation indicates
the other predictor variables included in the model.

The model estimated from the original dataset is given by
\[\widehat{Y} = b_{0\cdot1,2} + b_{1\cdot2} X_1 + b_{2\cdot1} X_2,\]
where \(b_{1\cdot2}\) is the sample coefficient on \(X_1\) given that
\(X_2\) is in the model, and \(b_{2\cdot1}\) is the sample coefficient
on \(X_2\) given that \(X_1\) is in the model.

An equivalent framing to the original data model is given for the
permuted data. For example, if \(Y\) is permuted to get \(Y^*,\) then
the model estimated from the permuted dataset is given by
\[\widehat{Y}^* = b^*_{0\cdot1,2} + b^*_{1\cdot2} X_1 + b^*_{2\cdot1} X_2,\]
where \(b^*_{1\cdot2}\) is the sample coefficient on \(X_1\) given that
\(X_2\) is in the model, and \(b^*_{2\cdot1}\) is the sample coefficient
on \(X_2\) given that \(X_1\) is in the model, while it is \(Y^*\) being
regressed on \(X_1\) and \(X_2.\)

All standard errors of the coefficients (\(SE(b)\) and \(SE(b^*)\)) are
calculated using the normal ordinary least squares (OLS) formula used in
standard linear regression software. Additionally, it is worth pointing
out that underlying all of the methods presented in Appendix
\ref{sec-appendix} are observations, which are assumed to be independent
of one another.

\subsection{Permutation algorithms}\label{sec-appendix-perm}

\subsubsection{\texorpdfstring{Permute
\(Y\)}{Permute Y}}\label{permute-y}

\paragraph*{\texorpdfstring{Summary of algorithm when permuting \(Y\)
(Manly
1986)}{Summary of algorithm when permuting Y (Manly 1986)}}\label{summary-of-algorithm-when-permuting-y-manly86}
\addcontentsline{toc}{paragraph}{Summary of algorithm when permuting
\(Y\) (Manly 1986)}

\begin{enumerate}
\def\labelenumi{\arabic{enumi}.}
\item
  Fit the original model and obtain coefficient estimates
  (\(b_{0\cdot 1,2},\) \(b_{1\cdot 2},\) and \(b_{2\cdot 1}\)) and
  corresponding standard error estimates (\(SE(b_{0\cdot 1,2}),\)
  \(SE(b_{1\cdot 2}),\) and \(SE(b_{2\cdot 1})\)):
  \[\widehat{Y} = b_{0\cdot1,2} + b_{1\cdot2}X_1 + b_{2\cdot1}X_2\]
\item
  Permute \(Y\) to obtain \(Y^*.\)
\item
  Fit a model, now including \(X_1\), on the permuted \(Y^*\) values to
  obtain permuted coefficient estimates (\(b^*_{0\cdot 1,2},\)
  \(b^*_{1\cdot 2},\) and \(b^*_{2\cdot 1}\)) and corresponding standard
  error estimates (\(SE(b^*_{0\cdot 1,2}),\) \(SE(b^*_{1\cdot 2}),\) and
  \(SE(b^*_{2\cdot 1})\)):
  \[\widehat{Y}^* = b^*_{0\cdot1,2} + b^*_{1\cdot2}X_1 + b^*_{2\cdot1}X_2\]
\item
  Repeat steps 2 and 3 \(P\) times. For example, \(P\) = 1000.
\item
  From the \(P\) copies of \(b^*_{1\cdot 2}\) and \(P\) copies of
  \(SE(b^*_{1\cdot 2}),\) calculate \(P\) copies of \(t^*\) to form the
  permuted null sampling distribution:
  \begin{equation}t^* = \frac{b_{1\cdot2}^* - 0}{SE(b_{1\cdot2}^*)} \label{t_y} \end{equation}
\item
  Compare the observed test statistic to the permuted null sampling
  distribution from step 5:
  \[t_{obs} = \frac{b_{1\cdot2} - 0}{SE(b_{1\cdot2})}\]
\end{enumerate}

\subsubsection{\texorpdfstring{Permute
\(X_1\)}{Permute X\_1}}\label{sec-alg-x1}

\paragraph*{\texorpdfstring{Summary of algorithm when permuting \(X_1\)
(Draper and Stoneman
1966)}{Summary of algorithm when permuting X\_1 (Draper and Stoneman 1966)}}\label{summary-of-algorithm-when-permuting-x_1-draper66}
\addcontentsline{toc}{paragraph}{Summary of algorithm when permuting
\(X_1\) (Draper and Stoneman 1966)}

\begin{enumerate}
\def\labelenumi{\arabic{enumi}.}
\item
  Fit the original model and obtain coefficient estimates and
  corresponding standard error estimates:
  \[\widehat{Y} = b_{0\cdot1,2} + b_{1\cdot2}X_1 + b_{2\cdot1}X_2\]
\item
  Permute \(X_1\) to obtain \(X^*_1.\)
\item
  Fit a model on the permuted \(X_1^*\) values to obtain permuted
  coefficient estimates and corresponding standard error estimates:
  \[\widehat{Y} = b^*_{0\cdot1,2} + b^*_{1\cdot2}X^*_1 + b^*_{2\cdot1}X_2\]
\item
  Repeat steps 2 and 3 \(P\) times.
\item
  From the \(P\) copies of \(b_{1\cdot2}^*\) and \(P\) copies of
  \(SE(b_{1\cdot2}^*),\) calculate \(P\) copies of \(t^*\) to form the
  permuted null sampling distribution:
  \begin{equation}t^* = \frac{b_{1\cdot2}^* - 0}{SE(b_{1\cdot2}^*)} \label{t_x1} \end{equation}
\item
  Compare the observed test statistic to the permuted null sampling
  distribution from step 5:
  \[t_{obs} = \frac{b_{1\cdot2} - 0}{SE(b_{1\cdot2})}\]
\end{enumerate}

\subsubsection{Permute reduced model residuals}\label{sec-alg-red}

\paragraph*{Summary of algorithm when permuting reduced model residuals
(Freedman and Lane
1983)}\label{summary-of-algorithm-when-permuting-reduced-model-residuals-freedmanlane83}
\addcontentsline{toc}{paragraph}{Summary of algorithm when permuting
reduced model residuals (Freedman and Lane 1983)}

\begin{enumerate}
\def\labelenumi{\arabic{enumi}.}
\item
  Fit the original model on \(X_2\) only and obtain coefficient
  estimates and corresponding standard error estimates of the reduced
  model: \[\widehat{Y} = b_{0\cdot2} + b_{2}X_2\]
\item
  Let the residuals be defined by
  \(R_{Y\cdot2} = Y - b_{0\cdot2} - b_{2}X_2,\) and permute
  \(R_{Y\cdot2}\) to obtain \(R^*_{Y\cdot2}.\) Define the permuted
  outcome variable as \(Y^* = b_{0\cdot2} + b_{2}X_2 + R^*_{Y\cdot2}.\)
\item
  Fit a model on the permuted \(Y^*\) values to obtain permuted
  coefficient estimates and corresponding standard error estimates:
  \[\widehat{Y}^* = b^*_{0\cdot1,2} + b^*_{1\cdot2}X_1 + b^*_{2\cdot1}X_2\]
\item
  Repeat steps 2 and 3 \(P\) times.
\item
  From the \(P\) copies of \(b_{1\cdot2}^*\) and \(P\) copies of
  \(SE(b_{1\cdot2}^*),\) calculate \(P\) copies of \(t^*\) to form the
  permuted null sampling distribution:
  \begin{equation}t^* = \frac{b_{1\cdot2}^* - 0}{SE(b_{1\cdot2}^*)} \label{t_red} \end{equation}
\item
  Compare the observed test statistic to the permuted null sampling
  distribution from step 5:
  \[t_{obs} = \frac{b_{1\cdot2} - 0}{SE(b_{1\cdot2})}\]
\end{enumerate}

It is not immediately obvious that a high correlation between \(X_1\)
and \(X_2\) leads to a non-null distribution. We provide evidence for
the claim in two ways: by describing the permutation scheme and by
giving a sketch for the proof of the claim.

\emph{Claim.} The correlation between \(Y^*\) and \(X_1\) is dependent
on the correlation between \(X_1\) and \(X_2.\)

\begin{enumerate}
\def\labelenumi{(\arabic{enumi})}
\item
  The permutation scheme is designed to create new values for the
  coefficient on \(X_1\) under the setting that \(X_1\) and \(Y\) are
  uncorrelated given \(X_2.\) Repeated sets of the ``permuted'' \(Y^*\)
  values are generated by adding noise (permuted residuals) to the
  fitted values from the model on \(X_2\) only. At first glance, it
  seems as though the method creates \(Y^*\) values that are associated
  with \(X_2\) (because they built from the \(X_2\)-model fitted values)
  and not associated with \(X_1\) (because they are built from a model
  that completely ignores \(X_1\)). However, if \(X_1\) and \(X_2\) are
  correlated, then \(Y^*\) values that are correlated with \(X_2\) will
  naturally be correlated with \(X_1.\)
\item
  In Appendix~\ref{sec-appendix-proof}, we give a proof sketch outlining
  the derivation for the covariance between \(Y^*\) and \(X_1.\)
  Although the proof is only outlined, the dependence on the correlation
  between \(X_1\) and \(X_2\) is clear. The approximate value of
  \(\rho(Y^*, X_1)\) derived in Appendix~\ref{sec-appendix-proof} is
  validated in Ye (2023).
\end{enumerate}

\subsubsection{Permute full model residuals}\label{sec-alg-full}

\paragraph*{\texorpdfstring{Summary of algorithm when permuting full
model residuals ({ter Braak}
1990)}{Summary of algorithm when permuting full model residuals (ter Braak 1990)}}\label{summary-of-algorithm-when-permuting-full-model-residuals-terbraak90}
\addcontentsline{toc}{paragraph}{Summary of algorithm when permuting
full model residuals ({ter Braak} 1990)}

\begin{enumerate}
\def\labelenumi{\arabic{enumi}.}
\item
  Fit the original model and obtain coefficient estimates and
  corresponding standard error estimates of the full model:
  \[\widehat{Y} = b_{0\cdot1,2} + b_{1\cdot2}X_1 + b_{2\cdot1}X_2\]
\item
  Let the residuals
  \(R_{Y\cdot1,2} = Y - b_{0\cdot1,2} - b_{1\cdot2}X_1 - b_{2\cdot1}X_2,\)
  and permute \(R_{Y\cdot1,2}\) to obtain \(R^*_{Y\cdot1,2}.\) Define
  the permuted outcome variable as
  \(Y^* = b_{0\cdot1,2} + b_{1\cdot2}X_1 + b_{2\cdot1}X_2 + R^*_{Y\cdot1,2}.\)
\item
  Fit a model on the permuted \(Y^*\) values to obtain permuted
  coefficient estimates and corresponding standard error estimates:
  \[\widehat{Y}^* = b^*_{0\cdot1,2} + b^*_{1\cdot2}X_1 + b^*_{2\cdot1}X_2\]
\item
  Repeat steps 2 and 3 \(P\) times.
\item
  From the \(P\) copies of \(b_{1\cdot2}^*\) and \(P\) copies of
  \(SE(b_{1\cdot2}^*),\) calculate \(P\) copies of \(t^*\) to form the
  permuted null sampling distribution:
  \begin{equation}t^* = \frac{b_{1\cdot2}^* - b_{1\cdot2}}{SE(b_{1\cdot2}^*)} \label{t_full} \end{equation}
\item
  Compare the observed test statistic to the permuted null sampling
  distribution from step 5:
  \[t_{obs} = \frac{b_{1\cdot2} - 0}{SE(b_{1\cdot2})}\]
\end{enumerate}

Note that the goal here is to understand the sampling distribution of
the statistic \(b^*_{1\cdot2}\) without specifying a null
hypothesis--exactly the work of bootstrapping (caveat: the work here is
done without replacement, where most bootstrapping is done with
replacement). Similar to the development of other bootstrapping methods
(Efron and Tibshirani 1994, 87), the underlying technical condition for
ter Braak's method is that: \begin{align}
F(x)_{\big(\frac{b_{1\cdot2}^* - b_{1\cdot2}}{SE(b_{1\cdot2}^*)}\big)} \approx F(x)_{\big(\frac{b_{1\cdot2} - \beta_{1\cdot2}}{SE(b_{1\cdot2})}\big)}, \label{eq:cdf-assump}
\end{align} where \(F\) is the cumulative distribution function of the
random variable described in the subscript of the function.

The sampling distribution of
\(t^* = \frac{b^{*}_{1\cdot2} - b_{1\cdot2}}{SE(b_{1\cdot2}^*)}\)
approximates the sampling distribution of
\(t = \frac{b_{1\cdot2} - \beta_{1\cdot2}}{SE(b_{1\cdot2})}\) due to the
theoretical underpinnings from bootstrapping. That is, the variability
of \(b_{1\cdot2}^*\) around \(b_{1\cdot2}\) mimics the variability of
\(b_{1\cdot2}\) around \(\beta_{1\cdot2}.\)

Under Equation (\ref{eq:cdf-assump}), we can use the \(t^*\)
distribution constructed from many permutations of the same dataset to
carry out a hypothesis test. If \(H_0\!\!: \beta_{1\cdot2} = 0\) is
true, then \(t_{obs} = \frac{b_{1\cdot2} - 0}{SE(b_{1\cdot2})}\) would
be a likely value in the \(t^*\) distribution, corresponding to a
non-significant \(p\)-value, resulting in a failure to reject \(H_0.\)

However, if \(H_A\!\!: \beta_{1\cdot2} \ne 0\) is true and, say,
\(\beta_{1\cdot2} = 47,\) then we would expect
\(\hat{t} = \frac{b_{1\cdot2} - 47}{SE(b_{1\cdot2})}\) to lie well
within the \(t^*\) distribution, while
\(t_{obs} = \frac{b_{1\cdot2} - 0}{SE(b_{1\cdot2})}\) would lie in the
margins, leading to a small \(p\)-value that concludes the test by
rejecting \(H_0.\)

Furthermore, {ter Braak} (1990) synthesizes equations from various
authors -- Efron (1982) for the bootstrap, along with Cox and Hinkley
(1974) and Lehmann and D'Abrera (1975) for the permutation -- to make
the following statements regarding the expected value and variance of
the estimated slope coefficients in the bootstrapping versus permutation
settings. \(b_{1\cdot2}^+\) corresponds to the coefficient derived from
a standard bootstrap (full model), while \(b_{1\cdot2}^*\) corresponds
to the permutation of the residuals under the full model. \begin{align}
E^{+}(b_{1\cdot2}^{+}) &= E^* (b_{1\cdot2}^*) = b_{1\cdot2} \label{eq:boot.1}\\
var^+(b_{1\cdot2}^+) &= (1 - 1/n)var^* (b_{1\cdot2}^*). \label{eq:boot.2}
\end{align}

Equation (\ref{eq:boot.1}) indicates that the expected values of
\(b_{1\cdot2}^+\) and \(b_{1\cdot2}^*\) are both \(b_{1\cdot2}.\)
Equation (\ref{eq:boot.2}) suggests that the variance of
\(b_{1\cdot2}^+,\) the bootstrapped estimate, is smaller than the
variance of \(b_{1\cdot2}^*.\) Hence, ter Braak uses Equations
(\ref{eq:boot.1}) and (\ref{eq:boot.2}), along with the order property
that \(b_{1\cdot2}^+\) and \(b_{1\cdot2}^*\) differ by \(O(1/n)\) in
second or higher order moments, to justify his proposal of the full
model residual permutation strategy.

\subsection{\texorpdfstring{Impact of correlation between \(X_1\) and
\(X_2\)}{Impact of correlation between X\_1 and X\_2}}\label{sec-appendix-proof}

\emph{Claim.} The correlation between \(Y^*\), derived from permuting
resdiuals from the reduced model, and \(X_1\) is dependent on the
correlation between \(X_1\) and \(X_2.\)

\emph{Proof sketch.} Throughout the proof sketch, there are places where
we have simplified the argument by considering statistics to be fixed
parameters (not an unreasonable approximation under large sample sizes
where Slutsky's Theorem holds (Slutsky 1925)). \begin{align}   
\rho(Y^*, X_1) &= \frac{cov(Y^*,X_1)}{\sqrt{var(Y^*)var(X_1)}} \notag
\end{align} Breaking down each part of the correlation between \(Y^*\)
and \(X_1,\) we compute both \(cov(Y^*, X_1)\) and \(var(Y^*).\)
\begin{align}
cov(Y^*, X_1) &= cov(b_{0\cdot2} + b_2X_2 + R^*_{Y\cdot2}, X_1) \notag \\
    &= cov(b_{0\cdot2},X_1)+cov(b_2X_2,X_1)+cov(R^*_{Y\cdot2},X_1)\notag\\
    &\approx cov(b_2 X_2, X_1)\label{eq:cov.1}\\
    &\approx b_2 \cdot cov(X_1, X_2) \label{eq:b1.1}\\
    &= b_2 \cdot \rho(X_1,X_2)\sqrt{var(X_1)var(X_2)}\notag
\end{align}

Note that \(b_{0\cdot2}\) and \(b_2\) are random variables because they
are statistics, so Equations (\ref{eq:b1.1},\ref{eq:b1.2},\ref{eq:b1.3})
are all approximate. In Equation (\ref{eq:cov.1}), we assume that
\(b_{0\cdot2}\) and \(R^*_{Y\cdot2}\) are both independent of \(X_1,\)
which leads to \(cov(b_{0\cdot2},X_1)=cov(R^*_{Y\cdot2}, X_1)=0.\)
\begin{align}
var(Y^*) &= var(b_{0\cdot2} + b_2 X_2 + R^*_{Y\cdot2})\notag\\
    &\approx (b_2)^2 var(X_2) + var(R^*_{Y\cdot2})\label{eq:b1.2}
\end{align}

\(var(R^*_{Y\cdot2})\) is broken down into pieces. \begin{align}
var(R^*_{Y\cdot2}) &= var(R_{Y\cdot2})\notag\\
    &= var(Y-b_{0\cdot2}-b_2X_2)\notag\\
    &\approx var(Y) + (b_2)^2 var(X_2) - 2b_2 cov(Y, X_2)\label{eq:b1.3}\\
cov(Y,X_2) &= cov(\beta_{0\cdot1,2}+\beta_{1\cdot2} X_1 + \beta_{2\cdot1} X_2 + \varepsilon, X_2)\notag\\
    &= \begin{aligned}[t]
        &{}cov(\beta_{0\cdot1,2}, X_2) + cov(\beta_{1\cdot2} X_1, X_2)\\ 
        &{}+ cov(\beta_{2\cdot1} X_2, X_2) + cov(\varepsilon,X_2)\\
    \end{aligned}\notag\\
    &= \beta_{2\cdot1} var(X_2) + \beta_{1\cdot2} cov(X_1, X_2) \label{eq:cov.2}
\end{align}

Plugging \(cov(Y,X_2)\) into \(var(R^*_{Y\cdot2})\) and
\(var(R^*_{Y\cdot2})\) into \(var(Y^*),\) we can approximate
\(var(Y^*).\) \begin{align}
var(Y^*) &\approx \begin{aligned}[t]
        &{}(b_2)^2 var(X_2) + var(Y) + (b_2)^2 var(X_2)\\
        &{} - 2b_2[\beta_{2\cdot1} var(X_2) + \beta_{1\cdot2} cov(X_1, X_2)] \\
    \end{aligned}\notag \\
    &= \begin{aligned}[t]
        &{}2(b_2)^2 var(X_2)+var(Y)\\
        &{} - 2b_2[\beta_{2\cdot1} var(X_2) + \beta_{1\cdot2} cov(X_1, X_2)]
    \end{aligned}\notag\\
    &= 2b_2(b_2-\beta_{2\cdot1})var(X_2) +var(Y) - 2 b_2\beta_{1\cdot2} cov(X_1,X_2)\notag\\
    &= 2b_2[(b_2-\beta_{2\cdot1})var(X_2)-\beta_{1\cdot2} cov(X_1,X_2)]+var(Y)
    \notag
\end{align} Putting it all together gives: \begin{align}
    \rho(Y^*, X_1) &= \frac{cov(Y^*,X_1)}{\sqrt{var(Y^*)var(X_1)}}\notag\\
    &\approx \frac{b_2 \cdot \rho(X_1,X_2)\sqrt{var(X_1)var(X_2)}}{\sqrt{var(Y^*)var(X_1)}}\notag\\
    &= \frac{b_2\cdot\rho(X_1,X_2)\sqrt{var(X_2)}}{\sqrt{var(Y^*)}}\notag\\
    &=  \frac{b_2\cdot\rho(X_1,X_2)\sqrt{var(X_2)}}{\sqrt{2b_2[(b_2-\beta_{2\cdot1})var(X_2)-\beta_{1\cdot2} cov(X_1,X_2)]+var(Y)}}.\label{eq:formulacorY*X2}
\end{align}

\end{document}